\documentclass[%
 reprint,
superscriptaddress,
 amsmath,amssymb,
 aps,prmaterials,
]{revtex4-2}

\usepackage{graphicx}
\usepackage{dcolumn}
\usepackage[dvipsnames]{xcolor}
\usepackage{bm}
\usepackage{braket}

\begin{document}
\title{Accurate Point Defect Energy Levels from Non-Empirical Screened Range-Separated Hybrid Functionals: the Case of Native Vacancies in ZnO}
\author{Sijia Ke}%
\affiliation{Department of Materials Science and Engineering, University of California at Berkeley, California 94720, USA}
\affiliation{Chemical Sciences Division, Lawrence Berkeley National Laboratory, Berkeley, California 94720, USA}
\author{Stephen E. Gant}%
\affiliation{Department of Physics, University of California at Berkeley, California 94720, USA}
\affiliation{Kavli Energy NanoSciences Institute at Berkeley, Berkeley, California 94720, USA}
\author{Leeor Kronik}%
\affiliation{Department of Molecular Chemistry and Materials Science, Weizmann Institute of Science, Rehovoth 76100, Israel}
\author{Jeffrey B. Neaton}%
\affiliation{Department of Physics, University of California at Berkeley, California 94720, USA}
\affiliation{Materials Sciences Division, Lawrence Berkeley National Laboratory, Berkeley, California 94720, USA}
\affiliation{Kavli Energy NanoSciences Institute at Berkeley, Berkeley, California 94720, USA}

\date{\today}

\begin{abstract}
We use density functional theory (DFT) with non-empirically tuned screened range-separated hybrid (SRSH) functionals to calculate the electronic properties of native zinc and oxygen vacancy point defects in ZnO, and we predict their defect levels for thermal and optical transitions in excellent agreement with available experiments and prior calculations that use empirical hybrid functionals. The ability of this non-empirical first-principles framework to accurately predict quantities of relevance to both bulk and defect level spectroscopy enables high-accuracy DFT calculations with non-empirical hybrid functionals for defect physics, at a reduced computational cost.
\end{abstract}

\maketitle

\section{Introduction}
Point defects play an essential role in the properties of semiconductors \cite{queisser1998}. They can function as a knob for tuning electronic transport properties or as an undesirable feature affecting semiconductor performance \cite{ohtomo1998,ryu2000,aoki2000,kim2003,mccluskey2020,desouza2023,wolfowicz2021,pastor2022}. To understand the effects of defects and better control them, a combination of experimental measurements and theoretical calculations are often required \cite{freysoldt2014}. While Kohn-Sham density functional theory (KS-DFT) \cite{hohenberg1964,kohn1965} is often the method of choice for such studies, commonly used local or semi-local exchange-correlation (xc) functionals tend to lead to delocalized orbitals, straining their ability to compute properties of localized defect states deep in the band gap \cite{kummel2008}. Along with underestimating the band gap, local or semi-local xc functionals tend to predict defect-induced bound mid-gap states as delocalized states in proximity with band edges \cite{gavartin2003,lany2008}. Therefore, to reliably compute and understand the electronic structure of defects, using more advanced techniques within large supercells containing one or more point defects is necessary. For example, ab initio many-body perturbation theory (e.g. the $GW$ approximation) can be employed \cite{lany2010,chen2015,chen2017,li2023,lewis2021,dernek2022}, although it imposes a high computational cost. 

Prior first-principles DFT calculations that have studied point defects in semiconductors or insulators have used either hybrid or DFT+U functionals \cite{alkauskas2008,finazzi2008,patterson2006,zhang2020,chen2013,chen2015,oba2008,lany2008,janotti2007,alkauskas2011,paudel2008,lyons2017,clark2010,wing2020a,lewis2017}. In recent studies, parameters of hybrid functionals or DFT+U have been empirically chosen to reproduce the experimentally-measured band gap of the pristine (defect-free) bulk system \cite{oba2008,lany2008,janotti2007,alkauskas2011,paudel2008,lyons2017,clark2010,wing2020a,lewis2017}. While such empirical approaches can be effective, they have limited predictive power for a few reasons. First, empirical tuning methods cannot be used to predict properties of materials prior to experimental study as they require knowledge of the measured band gaps. Furthermore, even when experimental band gap data are available, the impact of excitons, as well as thermal and zero-point renormalization of the measured band gap, must be accounted for. In practice, this can be challenging, introducing further error into the empirical tuning process. Second, even if the fundamental band gap is accurately captured using empirically tuned hybrid or DFT+U functionals, the calculated band edge values can vary relative to vacuum, affecting defect state energy relative to the band edge (relevant to the accuracy of defect transition levels) as well as other states in the supercell \cite{alkauskas2011,lany2008}. Third, methods reliant on empirical tuning often make the untested assumption that the functional parameters that produce a correct fundamental band gap in a pristine bulk unit cell can be predictive for supercells with defects.

Recently, a non-empirical Wannier-localized optimally tuned screened range separated hybrid (WOT-SRSH) functional has been developed \cite{wing2021} that predicts accurate band gaps for many bulk semiconductors and insulators \cite{ohad2023,ohad2022,sagredo2024} and serves as a high-quality starting point for $GW$-Bethe Salpeter equation \cite{ohad2023,gant2022,sagredo2024} and time-dependent DFT calculations \cite{ohad2023}. The success of WOT-SRSH can be attributed to two physically motivated constraints that the functional is non-empirically tuned to satisfy. First, it enforces the correct long-range asymptotic screening of the Coulomb interaction in solids; second, it satisfies an ansatz, first proposed by Ma and Wang \cite{ma2016}, whereby the ionization potential (IP) theorem \cite{perdew1982,almbladh1985} is enforced with removal of a maximally-localized Wannier function instead of the highest occupied orbital.

In this work, we use WOT-SRSH to compute the properties of two native ZnO point defects, namely the oxygen vacancy ($\text{V}_{\text{O}}$) and zinc vacancy ($\text{V}_{\text{Zn}}$), two well-studied defects of practical importance \cite{janotti2007,janotti2009,lyons2017}. Through the non-empirical tuning procedure of WOT-SRSH, we find the first two above-mentioned challenges associated with empirically-tuned DFT methods are largely remedied. Moreover, we find that the optimal parameters obtained from tuning WOT-SRSH in pristine bulk ZnO produce results that are very similar to those obtained from tuning in $\text{V}_{\text{O}}$-defective supercells, addressing the above-mentioned third limitation associated with empirically tuned functionals. Our example of oxygen vacancies in ZnO suggests a good transferability of tuned WOT-SRSH functional parameters between a defect-containing supercell and the pristine bulk limit and motivates future work on defect levels in other materials systems. We also note that while one-shot $GW$ calculations of the ZnO band gap have posed significant convergence difficulties \cite{rangel2020,shih2010,friedrich2011,stankovski2011}, Ohad et al. recently demonstrated \cite{ohad2023} that the standard application of the WOT-SRSH functional leads to an accurate band gap in ZnO, as we verify here. Moreover, our calculated thermal and optical transition levels using WOT-SRSH produce transition levels for $\text{V}_{\text{O}}$ and $\text{V}_{\text{Zn}}$ that agree well with prior experiments \cite{vlasenko2005,vlasenko2005a,laiho2008,evans2008,wang2009,knutsen2012} and prior calculations using empirical hybrid functionals \cite{lyons2017}. Our calculations suggest that WOT-SRSH can be predictive for defect properties in solids.

\section{Theory and Methods}
\subsection{Formalism}
For KS-DFT, even with the exact exchange-correlation (xc) functional, which is unknown, the fundamental band gap, defined as the difference between the ionization potential and the electron affinity, differs from the eigenvalue gap between the highest occupied and lowest unoccupied states by the exchange-correlation derivative discontinuity $\Delta_{\text{xc}}$ \cite{perdew1983,sham1983}. 
Additionally, in exact KS-DFT, the IP theorem \cite{perdew1982,almbladh1985} states that the highest occupied eigenvalue ($\epsilon_{\text{ho}}$) of an $N$-electron system is equal with the opposite sign to the ionization potential ($I$), 
\begin{equation}
\label{eq:IP_def}
    \epsilon_{\text{ho}} = E(N) - E(N-1) \equiv -I.
\end{equation}
Here, $E(N)$ is the total energy of the $N$ electron system and $E(N-1)$ is the total energy of the $N-1$ electron system. The IP theorem suggests that the error due to $\Delta_{\text{xc}}$ for the exact functional is only reflected in the difference between the electron affinity and energy of the lowest unoccupied orbital. However, for approximate DFT functionals, the missing xc derivative discontinuity often appears in approximately equal parts in both the highest occupied and lowest unoccupied eigenvalues \cite{allen2002,stein2012}. Hybrid functionals, which introduce a fraction of Fock (exact) exchange (EXX) into an xc functional, can reduce the needed derivative discontinuity, $\Delta_{\text{xc}}$, possibly driving it down to zero or even reversing its sign \cite{stein2012}. This potentially allows for better agreement with the IP theorem and, if so, typically results in a more accurate band gap \cite{kronik2012}.

Using the IP theorem as a tool for tuning hybrid functionals in practical calculations of crystalline solids is not straightforward, as the delocalized nature of Bloch states in periodic systems results in the IP theorem being trivially satisfied for any choice of functional \cite{mori-sanchez2008,kraisler2014,vlcek2015,gorling2015}. In order to address this issue, Ma and Wang \cite{ma2016} proposed an ansatz, referred to here as the Wannier-localized IP ansatz, that substitutes a maximally localized Wannier function (MLWF), $\phi$, representative of the highest energy valence band for the highest occupied KS orbital in the IP theorem so that
\begin{equation}
\label{eq:ma_wang_def}
    \braket{\phi|\hat{H}_{\text{DFT}}|\phi} = E(N) - E[\phi](N-1).
\end{equation}
The left-hand side of Eq. (\ref{eq:ma_wang_def}) is the expectation value of the energy of the MLWF in the DFT Hamiltonian. The localized nature of the Wannier function allows for a non-trivial tuning constraint for hybrid functionals \cite{wing2021}.

In a range-separated hybrid functional \cite{leininger1997} the Coulomb repulsion is partitioned, often in the form \cite{yanai2004} 
\begin{equation}
\label{eq:srsh_def}
    \frac{1}{r}=\frac{\alpha+\beta \text{erf}(\gamma r)}{r}+\frac{1-[\alpha+\beta \text{erf}(\gamma r)]}{r}.
\end{equation}
The exchange associated with the first term is treated exactly, while the exchange associated with the second term is approximated by semi-local exchange. Additionally, the correlation is handled entirely at the semi-local level. As indicated in Eq. (\ref{eq:srsh_def}), the RSH partition introduces three parameters: $\alpha$, $\beta$, and $\gamma$. $\alpha$ represents the fraction of EXX in the short range, $\alpha + \beta$ is the amount of long-range EXX, and $\gamma$ is the range-separation parameter that mediates the transition between the two limits. 

In this work, we use the SRSH framework \cite{refaely-abramson2013}, in which the amount of long-range EXX is chosen to enforce the correct long-range asymptotic behavior of the Coulomb potential, i.e., $\alpha + \beta = \varepsilon_{\infty}^{-1}$, where $\varepsilon_{\infty}$ is the orientationally averaged clamped-ion electronic dielectric constant. Finally, in Wannier-localized optimal tuning (WOT), the choice of $\{\alpha,\gamma\}$ is constrained to fulfill the Wannier-localized IP ansatz of Eq. (\ref{eq:ma_wang_def}) \cite{wing2021}, i.e., we seek $\{\alpha,\gamma\}$ such that 
\begin{eqnarray}\label{eq:deltai}
   \Delta I^{\{\alpha,\gamma\}}_{\text{MLWF}} = E^{\{\alpha,\gamma\}}_{\text{constrained}}[\phi](N-1) - E^{\{\alpha,\gamma\}}(N) \nonumber \\
    + \langle\phi| \hat{H}^{\{\alpha,\gamma\}}_{\text{SRSH}}|\phi\rangle
\end{eqnarray}
is zero. The first term on the right-hand side of Eq.\ (\ref{eq:deltai}) is the total energy of the $N-1$ electron system, calculated via a constrained DFT approach that depopulates the state $\phi$ by adding an energy penalty to its overlap with the eigenstates of the $N-1$ electron system ($\psi_i^{N-1}$, where $i$ is the eigenstate index) \cite{wing2021}. A correction due to the interaction between the induced positive charge associated with the Wannier function, its periodic images, and a compensating background charge \cite{wing2021,makov1995} needs to be included in this first term. The second term on the right-hand side of Eq. (\ref{eq:deltai}) is the total energy of the $N$ electron system, and the third term is the expectation value of the energy of the MLWF in the SRSH Hamiltonian, where typically the MLWF with the highest expectation value is used. More detailed information on the WOT-SRSH tuning procedure can be found in Ref. \cite{wing2021}.

In the case of a deep defect that introduces an occupied localized state with an energy level close to mid-gap, the localized nature of this defect state can in principle restore the non-triviality of satisfying Eq. (\ref{eq:IP_def}) \cite{miceli2018,bischoff2019a,bischoff2019,falletta2020,yang2022}. Here, we can leverage the neutral $\text{V}_{\text{O}}$ defect state in ZnO, which satisfies this condition, to utilize a different variation of $\Delta I$ based on Eq. (\ref{eq:IP_def}) for the non-pristine supercell that contains the defect state, namely,
\begin{eqnarray}\label{eq:deltai-def}
   \Delta I^{\{\alpha,\gamma\}}_{\text{DS}} = E^{\{\alpha,\gamma\}}(N-1) - E^{\{\alpha,\gamma\}}(N) \nonumber \\ 
    + \epsilon^{\{\alpha,\gamma\}}_{N}.
\end{eqnarray}
In Eq. (\ref{eq:deltai-def}), the subscript 'DS' stands for defect state. The first term of the equation is obtained by performing an $N-1$ electron calculation (again with charge corrections included). The third term, $\epsilon_{N}$, is the highest occupied eigenvalue of the $N$ electron system, which would correspond to the defect level.

\subsection{Computational details}
\label{subsec:calc_details}
In this work, we obtain and use non-empirically tuned optimal SRSH parameters using three approaches. The first two approaches both use WOT-SRSH with Eq.\ (\ref{eq:deltai}), once with a supercell of pristine bulk ZnO to obtain $\{\alpha, \gamma\}_{\text{MLWF},\text{pris}}$ and once with the $\text{V}_{\text{O}}$ containing supercell to obtain $\{\alpha, \gamma\}_{\text{MLWF},\text{V}_\text{O}}$. In the third approach, we use the defect-containing supercell and $\Delta I_{\text{DS}} = 0$ from Eq. (\ref{eq:deltai-def}) to obtain $\{\alpha, \gamma\}_{\text{DS},\text{V}_\text{O}}$. In the case of $\{\alpha, \gamma\}_{\text{MLWF},\text{V}_\text{O}}$, the MLWF is obtained from a wannierization procedure that allows for mixing with non-defect valence states (see SI \cite{si} for details). Had the wannierization procedure been carried out on just the single isolated defect band, the MLWF would have been exactly the defect wavefunction, leading to nearly the same ionization potential as $I_{ \text{DS}}$ (when both are performed with the same reciprocal space sampling). For all three approaches, we use $\alpha+\beta =\varepsilon_\infty^{-1}=1/3.63=0.275$, where the orientationally averaged clamped-ion electronic dielectric constant $\varepsilon_\infty$ for bulk ZnO is calculated using HSE06 \cite{heyd2006}, noting ZnO exhibits only mild anisotropy (see SI \cite{si} for details). These approaches result in different optimal $\{\alpha, \gamma\}$ pairs, as discussed in detail below.

For a given choice of tuning criteria, either Eq. (\ref{eq:deltai}) or Eq. (\ref{eq:deltai-def}), $\Delta I$ changes monotonically from $\Delta I (\alpha,\gamma = 0)$ and asymptotically approaches $\Delta I (\alpha = \varepsilon_\infty^{-1})$ as $\gamma$ increases for a fixed $\alpha$. To find $\gamma$ that gives $\Delta I (\alpha,\gamma) =0$, $\Delta I (\alpha,\gamma = 0)$ and $\Delta I (\alpha = \varepsilon_\infty^{-1})$ are required to have opposite signs, restricting the subspace of optimal $\alpha$, i.e., only one $\gamma$ can lead to $\Delta I (\alpha,\gamma) =0$ with a given $\alpha$ that belongs to a particular range. There can be multiple optimal $\{\alpha,\gamma\}$ pairs for each of the tuning criteria, but the effect of these different pairs of parameters on band gap values is relatively modest, on the order of 100 meV at the most and 10 meV more typically \cite{wing2021,gant2022}. As different choices of tuning criteria (and atomic distortions) lead to different ranges for $\alpha$, in this work, we examine in particular the $\gamma$ that satisfies $\Delta I =0$ with a fixed $\alpha$ set to be 0.2 for tuning with a MLWF in a supercell containing one $\text{V}_{\text{O}}$, and set to 0.35 or $\varepsilon_\infty^{-1}$ for tuning with defect eigenstate (DS) in a supercell containing one $\text{V}_{\text{O}}$ and tuning with a MLWF in a pristine bulk supercell (see Results section below). In the case of $\alpha =\varepsilon_\infty^{-1}$, the functional reduces to a global hybrid one that is independent of $\gamma$. 

Depending on the supercell size and the atomic character of the defect states, the defect wavefunction can overlap and interact with its images due to periodic boundary conditions, leading to shifts in its eigenvalues relative to the dilute limit. To account for this in calculating $\Delta I_{\text{DS}}$, we use a $2\times2\times2$ k-mesh and average the eigenvalues of the defect band when computing the third term on the right-hand side of Eq. (\ref{eq:deltai-def}). In contrast, for $\Delta I_{\text{MLWF}}$ we only sample the $\Gamma$ point in reciprocal space. Supercell size and reciprocal space convergence details are given in the SI \cite{si}.

Local atomic scale structural distortions, introduced by a point defect, can change the localization and energy of the defect state, impacting optimal tuning. We therefore consider structural distortions around the point defect. Details of these distortions are given in the SI \cite{si}. The defect formation energy and thermal charge transition level are calculated using the scheme presented in Refs.\ \cite{freysoldt2014,lyons2017}. Optical (vertical) transitions are calculated using the Franck-Condon principle, neglecting excitonic effects, as is standard \cite{freysoldt2014}. The vertical charge transition level is defined as \begin{equation}
\label{eq:vert_tran}
    \Delta E(q_{i}|q_{f}) = E(q_{f};R) + (q_{f}-q_{i}) \epsilon_{CBM/VBM} - E(q_{i};R),
\end{equation}
where $q_i$ and $q_f$ are the initial and final charge states of the defect, $E(q;R)$ is the total energy (with image charge corrections \cite{freysoldt2009,kumagai2014a,gake2020} included), $\epsilon_{CBM/VBM}$ is the conduction band minimum or the valence band maximum eigenvalue, depending on which is involved, and $R$ represents the structural configuration. For example, the vertical transition level (0/1+), associated with electron capture by $\text{V}^{1+}_{\text{O}}$ is calculated as 
\begin{eqnarray}\label{eq:vctl}
    \Delta E(0|1+) = E(1+,R_{0}) + \epsilon_{CBM} - E(0,R_0).
\end{eqnarray}

Finally, we note that we use the charge corrections \cite{freysoldt2009,kumagai2014a,gake2020} that go beyond the simple point charge assumption \cite{makov1995} for tuning $\Delta I_{\text{DS}}$ or for computing the transition level of $\text{V}_{\text{O}}$. These higher-order non-point-charge corrections can be on the order of 10 meV for the defect structures studied here. Incorporating them is consistent with best practice tuning of finding the root of $\Delta I = 0$ to within 20 meV. 
\section{Results}
\subsection{Parameter tuning}
\label{subsec:param_sel}
Using the tuning approaches outlined in Section \ref{subsec:calc_details}, we compute the three sets of optimally-tuned pairs: $\{\alpha,\gamma\}_{\text{MLWF},\text{pris}}$ (using a MLWF in pristine bulk ZnO), $\{\alpha, \gamma\}_{ \text{MLWF},\text{V}_\text{O}}$ (using a MLWF in a supercell containing one $\text{V}_{\text{O}}$), and $\{\alpha,\gamma\}_{\text{DS},\text{V}_\text{O}}$ (using a defect eigenstate in a supercell containing one $\text{V}_{\text{O}}$). We consider supercells containing $\text{V}_{\text{O}}$ with and without structural relaxation, to examine the sensitivity of the optimal tuning to the local geometry around the defect. In the case of a neutral $\text{V}_{\text{O}}$ in ZnO, the structural relaxation further localizes the defect state, as shown in the SI, Figure S1 \cite{si}. Tuning with a zinc vacancy ($\text{V}_{\text{Zn}}$) is beyond the scope of this work as its defect state is more delocalized than $\text{V}_{\text{O}}$, with the defect wavefunction extending to its four oxygen neighbors \cite{lany2010,lyons2017}, requiring a larger supercell at greater computational cost.

In the following, we use an experimental reference ZnO fundamental band gap value ($E_g^{ref}$) of 3.60 eV. This value is obtained by removing exciton binding energy \cite{mang1995} and phonon contributions \cite{manjon2003,tsoi2006,miglio2020} from the experimentally known band gap (see the SI \cite{si} for details). 
Figure 1 shows a contour plot of relative band gap values, $\Delta E_g^{\alpha,\gamma}=E_g^{\alpha,\gamma}-E_g^{ref}$, where the gap $E_g^{\alpha,\gamma}$ is computed using a primitive unit cell of bulk ZnO, as a function of the SRSH parameters $\{ \alpha,\gamma \}$, treated as free parameters. Figure 1 also shows, as contour dashed lines, $\Delta E_g^{\alpha,\gamma}$ values obtained from optimally-tuned $\{\alpha,\gamma\}$ pairs, tabulated values in Table \ref{tbl:alphagamma}. Finally, for the five tuning scenarios studied, we show (in symbols) one specific $\{\alpha,\gamma\}$ pair, selected based on typical default values of ${\alpha}$=0.2 for contour dashed lines lying below $\alpha = \varepsilon_{\infty}^{-1}$, ${\alpha}$=0.35 for contour dashed lines lying above $\alpha = \varepsilon_{\infty}^{-1}$

\begin{table}[hbt!]\centering
\caption{Tuned $\{\alpha, \gamma\}$ pairs obtained with different methods. DS indicates that it is tuned using the isolated defect eigenstate. When $\alpha = \varepsilon_\infty^{-1} = 0.275$, our hybrid functional is now global and no longer range-separated, meaning that any value for $\gamma$ gives the same result since the short and long range regimes are identically screened. $\Delta I$ refers to the equation where we tune $\{\alpha, \gamma\}$ for $\Delta I=0$: 'MLWF' uses Eq. \ref{eq:deltai}, 'DS' uses Eq. \ref{eq:deltai-def}.}
\begin{tabular}{c|l|ll|l}
\hline
\multicolumn{1}{l|}{Structure} & $\Delta I$  & $\alpha$ & $\gamma$ / $\text{\AA}^{-1}$&Band gap / eV\\ \hline
\multicolumn{1}{l|}{pristine}  & MLWF& 0.35  & 0.47   &3.80\\ \hline
{$\text{V}_{\text{O}}$-relaxed}    & MLWF& 0.2   & 0.26   &2.94\\
                               & DS& 0.35  & 0.4    &3.84\\ \hline
{$\text{V}_{\text{O}}$-unrelaxed}     & MLWF& 0.2   & 0.2    &2.90\\
                               & DS& 0.275 & na     &3.44\\ \hline
\end{tabular}
\label{tbl:alphagamma}
\end{table}

\begin{figure}[htbp!]
    \centering
     \includegraphics[width=0.48\textwidth]{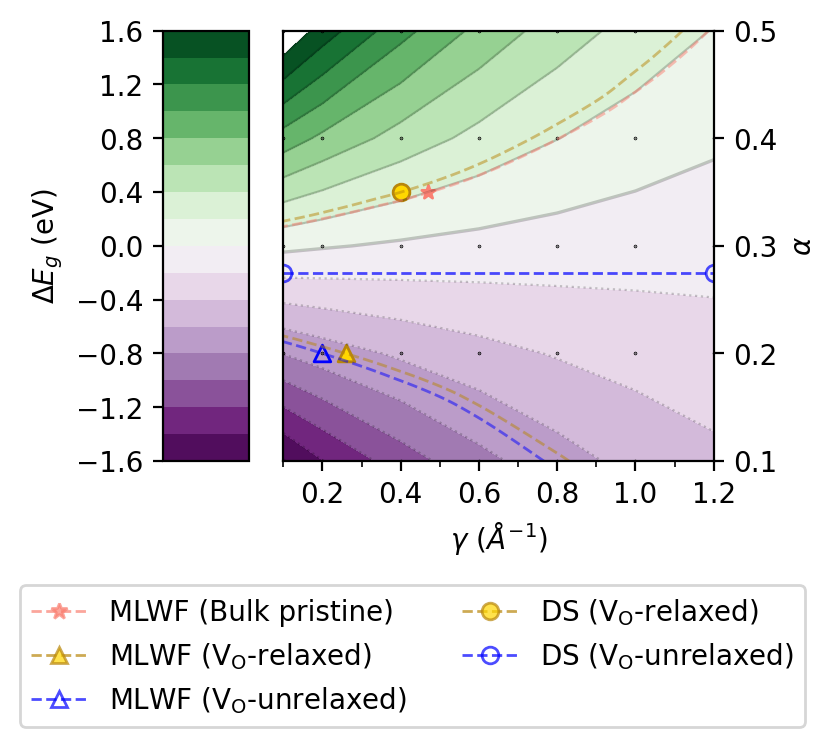}
    \caption{Fundamental band gap contours, calculated with WOT-SRSH using different $\{\alpha,\gamma\}$ pairs in bulk ZnO, shown as deviations from a reference band gap value of 3.6 eV. Dashed lines indicate curves corresponding to optimally-tuned $\{\alpha, \gamma\}$ pairs obtained from different methods, as given in the legend. Symbols indicate a specific $\{\alpha, \gamma\}$ pair from the dashed lines, obtained for $\alpha=0.2$ for lines with $\alpha$ below $\varepsilon_{\infty}^{-1}$ and $\alpha=0.35$ for lines with $\alpha$ above $\varepsilon_{\infty}^{-1}$. For one line, $\alpha = \varepsilon_{\infty}^{-1}$, the result is $\gamma$-independent. 'pris' indicates tuning in a pristine bulk cell; '$\text{V}_{\text{O}}$' indicates tuning in a supercell with an oxygen vacancy. 'MLWF' indicates using the Wannier-localized IP ansatz and 'DS' indicates using the IP theorem with charge removed from the defect states.}
    \label{fig:eg}
\end{figure}

The optimally tuned parameters of pristine bulk ZnO (Fig. \ref{fig:eg}, red star), $\{\alpha, \gamma\}_{\text{MLWF}, \text{pris}} = \{0.35,0.47 \text{\AA}^{-1}\}$ result in a fundamental bulk band gap of 3.80 eV, which is larger than the reference experimental value by 0.2 eV. We note that these values slightly differ from those previously reported for ZnO using the same method \cite{ohad2023}, namely a gap of 3.53 eV based on optimal SRSH parameters of $\{\alpha, \gamma, \varepsilon_\infty\} = \{0.30, 1.30 \text{\AA}^{-1}, 3.57\}$. This is because the prior work iterated the tuning procedure to self-consistency, in the sense of a choice of $\{\alpha, \gamma, \varepsilon_\infty\}$ that leads to the same averaged macroscopic dielectric constant, $\varepsilon_\infty\}$. Here we instead simply use the value of $\varepsilon_\infty$ calculated from HSE06 (more details in SI \cite{si}). It is readily observed that the differences owing to this "one-shot" $\varepsilon_\infty$ approach are not negligible but are still rather small. 
However, regarding $\varepsilon_\infty$, we note that for materials with a high anisotropy or a functional-sensitive dielectric environment, using a one-shot calculation of the orientationally averaged clamped-ion electronic dielectric constant may lead to errors. Nevertheless, the WOT-SRSH functional has been accurately applied in highly anisotropic van der Waals materials \cite{camarasa-gomez_transferable_2023,camarasa-gomez2024}.

The $\{\alpha, \gamma\}_{\text{DS}, \text{V}_\text{O}}$ (Fig. \ref{fig:eg}, circle) values obtained for relaxed and unrelaxed defect structures lead to fundamental band gaps of 3.84 eV and 3.44 eV respectively, which are within 0.25 eV of the reference fundamental band gap. $\{\alpha, \gamma\}_{\text{MLWF},\text{V}_\text{O}}$ (Fig. \ref{fig:eg}, triangles) for relaxed and unrelaxed defect structures yield fundamental band gaps of 2.94 eV and 2.90 eV, respectively, which are about 0.6 eV lower than the reference value. Among these different approaches, using $\Delta I_{\text{MLWF}}=0$ as a tuning criterion in pristine ZnO and using $\Delta I_{\text{DS}}=0$ in a supercell containing a $\text{V}_{\text{O}}$ produce the best agreement with the measured reference gap. The accuracy of tuning with $\Delta I_{\text{MLWF}}$ in a pristine cell is consistent with previous WOT-SRSH studies \cite{ohad2023}, and the accuracy of using defects to obtain hybrid functional parameters for the fundamental band gap has also been reported in the past \cite{miceli2018,bischoff2019,bischoff2019a,yang2022}. We attribute the relative inaccuracy of using $\Delta I_{\text{MLWF}}$ with ${\text{V}_{\text{O}}}$ defective supercell to the selected Wannier function being a mixture of the defect state and other bulk valence bands. If the wannierization manifold is limited to only the isolated defect state, then the Wannier function is essentially the defect state orbital and nearly identical to $\{\alpha,\gamma\}_{\text{DS},\text{V}_{\text{O}}}$ (with only $\Gamma$ point sampled in reciprocal space). This result illustrates the potential risk of constructing a Wannier function that overly dilutes the character of the topmost occupied state.

Although the IP theorem only guarantees that the eigenvalue of the highest occupied orbital has direct physical meaning \cite{perdew1982}, in prior work we have found that WOT-SRSH eigenvalues as much as a few eV away from the valence band edge can still provide good approximations to quasiparticle excitation energies \cite{refaely-abramson2013,kronik2014,wing2021,gant2022,ohad2022,ohad2023}. Specifically for $\text{V}_{\text{O}}$-containing ZnO, with a localized highest occupied defect state, this implies that other states, including the bulk valence band maximum, can be well described. Perhaps more strikingly, the results obtained from $\{\alpha, \gamma\}_{\text{MLWF},\text{pris}}$ and $\{\alpha,\gamma\}_{\text{DS},\text{V}_{\text{O}}}$ for the relaxed structure only differ by 0.04 eV, showing that WOT-SRSH parameters obtained in the pristine cell can be directly used in the defect-containing supercell with little deviation from the IP theorem. In other words, we find that WOT-SRSH parameters are \textit{transferable} between pristine and defective supercells in ZnO. This suggests that one can simply use the pristine tuned parameters, saving the computational cost of tuning in a large supercell with defects.
On the contrary, tuning with unrelaxed structures leads to band gap values different from those obtained for the relaxed structure, by as much 0.4 eV, which is likely related to a different degree of localization induced by the distortion of the atomic geometry near the defect. Therefore, the role of relaxation in determining optimally-tuned hybrid functional parameters should be examined. 

We find that the sensitivity of the defect state eigenvalue to the hybrid functional parameters is generally smaller than that of the valence band maximum, which means that the computational uncertainty in the band gap is larger than the uncertainty in the defect state energy. This implies there can be a larger uncertainty in the band gap when using defects to enforce IP theorem. An example of this can be seen in the SI Figure S6 \cite{si}, where the sensitivity of the valence band maximum and the defect state to $\{\alpha,\gamma\}$ are shown. Using a defect state with minimal wavefunction spread (which reduces its interaction with its periodic images) and with its defect state energy resonant with the valence band maximum can reduce this sensitivity. We also refer to past studies \cite{miceli2018,bischoff2019,bischoff2019a,yang2022} for the sensitivity of tuning with different types of defects to obtain hybrid functional parameters for the fundamental band gap.\\
\subsection{Oxygen vacancy}
\label{subsec:VO}
\begin{figure}[htbp!]
    \centering
    \includegraphics[width=0.48\textwidth]{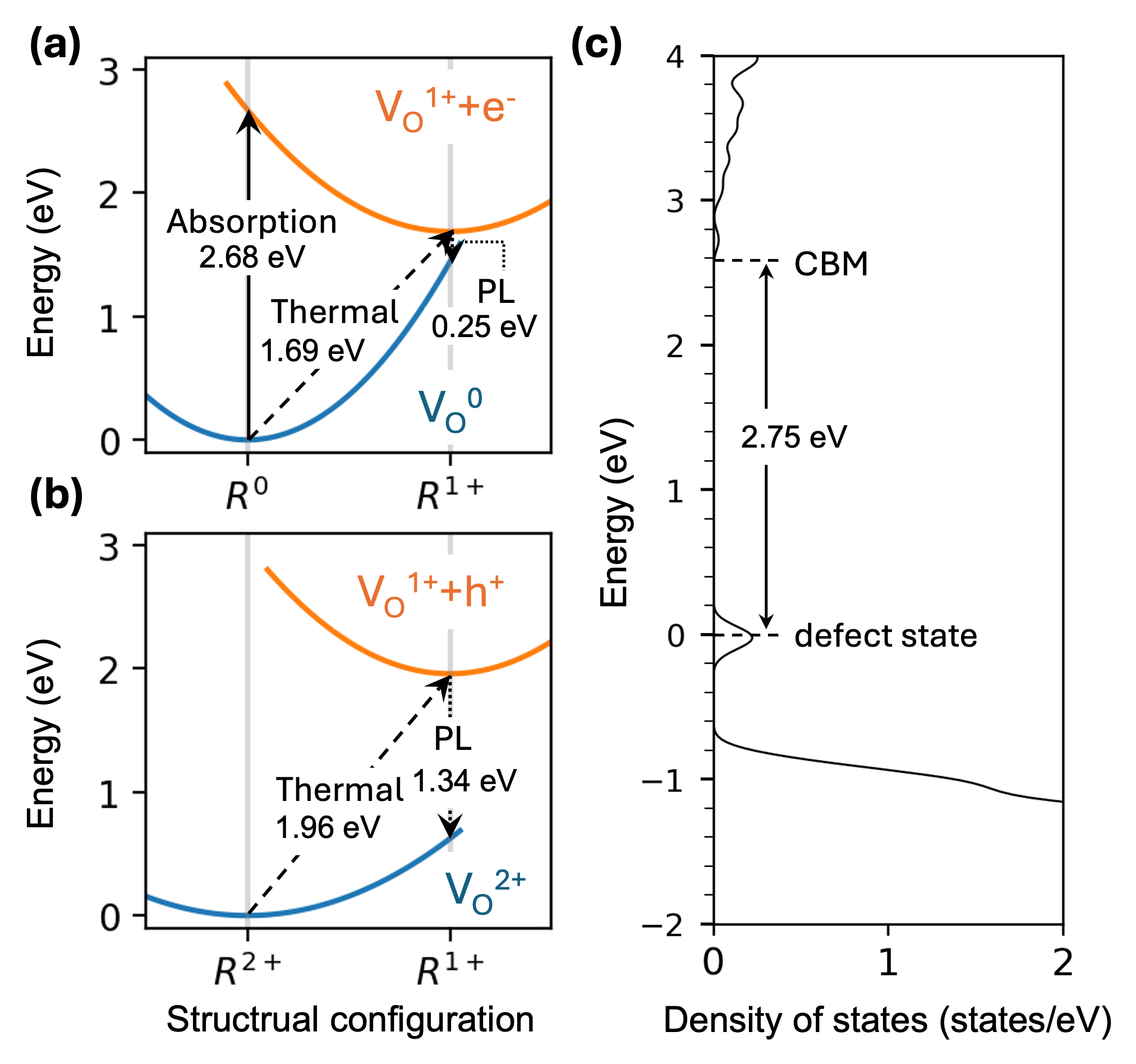}
    \caption{(a,b) Schematic configuration-coordinate diagrams with thermal (dashed arrow), optical absorption (solid arrow), and photoluminescence (PL) (dotted arrow) charge transitions in ZnO, with (a) and (b) corresponding to electron and hole capture by an oxygen vacancy in the 1+ charge state. The ground and excited states are given in blue and orange lines, respectively.(c) Density of states obtained from the neutral oxygen vacancy-containing ZnO supercell, with the energy position of the defect states indicated.}
    \label{fig:conf}
\end{figure}

Using the optimally-tuned parameters $\alpha = \varepsilon_\infty^{-1} = 0.275$, we calculate the charge transition levels for the oxygen vacancy, $\text{V}_{\text{O}}$. Calculated formation energies (see SI Figure S7 \cite{si}) indicate that only the neutral and 2+ charge states are stable, in agreement with prior calculations \cite{lyons2017}. In Figure \ref{fig:conf}, we plot schematic configuration-coordinate diagrams with the two aforementioned stable charges being the initial ground state, along with the density of states (DOS) for neutral $\text{V}_{\text{O}}$, obtained with WOT-SRSH. We also provide the corresponding transition energies in Table \ref{tab:trans}, where it is compared to prior literature \cite{lyons2017} and experimental electron paramagnetic resonance (EPR) photoabsorption measurements \cite{vlasenko2005,vlasenko2005a,evans2008,laiho2008}; other possible charge transition values are listed in SI Table S2 \cite{si}. We use the standard notation for a transition of (initial charge state/final charge state). The thermal and vertical transition levels (0/1+) of 1.69 eV and 2.68 eV, respectively, are close to the signal threshold (1.75 eV) and the peak of EPR (2.5 eV) under photo-absorption in electron-irradiated ZnO samples \cite{vlasenko2005a,vlasenko2005,evans2008,laiho2008}. The photoluminescence (PL) peak associated with the (1+/0) vertical transition is 0.25 eV, a value that has been too small to detect in the prior experiments. 

The computed thermal (2+/1+) transition level is 1.96 eV, close to the EPR signal threshold of 2.1 eV observed in gamma-ray-irradiated and as-grown ZnO samples \cite{evans2008,laiho2008a,wang2009}. The vertical transition level (2+/1+) cannot be computed in this work due to the unoccupied defect state associated with $\text{V}_{\text{O}}^{2+}$ merging into conduction bands in our calculations, as also found in prior work \cite{lany2010}. The photoluminescence peak associated with the (2+/1+) vertical transition is 1.34 eV, at the edge of the PL-EPR detection range \cite{vlasenko2005,vlasenko2005a}. Based on a comparison with experiments, differently synthesized ZnO samples may have different oxygen vacancy charge states, where electron-irradiated samples contain mostly neutral oxygen vacancies and gamma-irradiated and as-grown samples contain mostly 2+ oxygen vacancies. Note that these calculated transition energies do not support the assignment of $\text{V}_{\text{O}}$ as the origin of green luminescence in ZnO \cite{janotti2009,ye2005}, a phenomenon whose origin has been long debated \cite{lyons2017}. 

As can be seen in Table \ref{tab:trans}, the majority of our results are consistent with prior experiments and prior theoretical work using empirical hybrid functional approaches. However, the vertical charge transitions $(1+,R^{1+}/0,R^{1+})$ and $(1+,R^{1+}/2+,R^{1+})$ show differences compared to previous calculations with empirical hybrid functionals \cite{lyons2017,freysoldt2009}. These differences likely originate from the different charge screening schemes used to compute the vertical transition. Our work utilizes a recently-updated scheme \cite{gake2020} where the interaction of an added charge ($q_{f}-q_{i}$) and its periodic images is screened by electrons, and the interaction of added charge and initial charge is screened by both electrons and ions; past studies \cite{lyons2017} used a charge screening scheme with the same dielectric tensor. 

\begin{table*}[htbp!]
\label{tab:trans}
\caption{Calculated thermal and optical transitions (absorption and PL) associated with $\text{V}_{\text{O}}$ electron capture and $\text{V}_{\text{O}}^{2+}$ hole capture, compared to experimental EPR signal thresholds and peak positions for $\text{V}_{\text{O}}$ from the literature. We note the initial and final charge state and structure in the brackets, where the superscript $x$ of $R^{x}$ represents the atomic structure that is internally relaxed under a charge state of $x$. We compare the threshold of EPR to the thermal transition and the peak position of the EPR to the relevant optical absorption transition.}
\begin{tabular}{lllllll}
\hline
\multicolumn{7}{c}{Calculation} \\ \hline
 \multicolumn{1}{l|}{}& \multicolumn{3}{c|}{$\text{V}_{\text{O}}$ electron capture}& \multicolumn{2}{c|}{$\text{V}_{\text{O}}^{2+}$ hole capture}&\\ 
 \multicolumn{1}{l|}{}&Thermal&Absorption&\multicolumn{1}{l|}{PL}&Thermal&\multicolumn{1}{l|}{PL}&Thermal\\ 
 \multicolumn{1}{l|}{}&$(0,R^0/1+,R^{1+})$&$(0,R^0/1+,R^{0})$&\multicolumn{1}{l|}{$(1+,R^{1+}/0,R^{1+})$}&$(2+,R^{2+}/1+,R^{1+})$&\multicolumn{1}{l|}{$(1+,R^{1+}/2+,R^{1+})$}&$(0,R^{0}/2+,R^{2+})$\\ \hline
\multicolumn{1}{l|}{Our work}   & 1.69 & 2.68 & \multicolumn{1}{l|}{0.25} & 1.96 & \multicolumn{1}{l|}{1.34} & 1.85 \\ \hline
\multicolumn{1}{l|}{Ref \cite{lyons2017} }     & 1.68 & 2.67 & \multicolumn{1}{l|}{0.62} & 2.18 & \multicolumn{1}{l|}{0.91}& 1.93 \\ 
\end{tabular}
\setlength{\tabcolsep}{18pt}
\begin{tabular}{lllllll}
\hline
\multicolumn{7}{c}{Experiment}\\ \hline
Sample & Threshold & Peak & & Sample & Threshold  & Peak  \\ \cline{1-3} \cline{5-7} 
$\text{e}^{-}$-irriated \cite{vlasenko2005} & 1.76 & 2.48 & & $\gamma$-irriated \cite{laiho2008} & 2.1  & 2.5  \\
$\text{e}^{-}$-irriated \cite{laiho2008} & 1.75 & 2.5 & & as-grown \cite{wang2009} & 2.1  & 2.6  \\ \hline
\end{tabular}
\end{table*}

Generally speaking, the vertical charge transition level can be calculated from the ground state energy difference corresponding to the same structure in two different charge states, or directly from the quasiparticle eigenvalue difference of one charge state. The first approach is by far more widespread, because ground-state energy differences are considerably less sensitive to the functional used. We therefore point out the similarity between results we obtain from Eq. (\ref{eq:deltai-def}) for $\Delta I$ in optimal tuning and Eq. (\ref{eq:vctl}) for the optical charge transition level (0/+1). We can combine these two equations and write $\Delta E(0|+1) \approx \Delta I + \epsilon_{CBM} - \epsilon_{def}$. With a small $\Delta I$, a vertical transition level (0/1+) can be directly obtained from the eigenvalues from one initial neutral ground state calculation, a significant saving of computational time. The density of states in Fig. \ref{fig:conf} illustrates the energy difference of 2.75 eV between the averaged defect state eigenvalues and conduction band minimum, close to the calculated vertical charge transition level of 2.68 eV, suggesting that our non-empirically tuned SRSH parameters lead to physical results. However, for systems with an initially charged defect, computed states associated with the charged defect require an energy correction due to electrostatic interactions between periodic images \cite{chen2015}. The non-trivially mixed screening scheme mentioned before prevents such straightforward calculations from the density of states and requires more complex analysis \cite{gake2020,falletta2020}.\\

\subsection{Zinc vacancy}
Using the same optimally-tuned hybrid parameters as in \ref{subsec:VO}, we further study the defect transition levels of the Zn vacancy, $\text{V}_{\text{Zn}}$. First, we compute the relative formation energies for 5 charge states, (2-, 1-, 0, 1+, 2+), and find they can all become stable depending on the position of the Fermi level; see SI Figure S7 \cite{si} for details. All optical and thermal transitions arising from these possibilities are tabulated in SI Table S3 and compared with prior reports \cite{si}. All calculated charge states of $\text{V}_{\text{Zn}}$ have unpaired spins, except for $\text{V}_{\text{Zn}}^{2-}$. We find that the charges are not localized on the center of Zn vacancies but rather on the neighboring oxygen atoms. This is consistent with prior computational studies using the $GW$ approximation or empirical hybrid functionals \cite{lany2010,clark2010,lyons2017}. Ensuring an accurate charge correction with a complex charge distribution is not trivial, and here we include a point charge correction for $\text{V}_{\text{Zn}}$ (see details in SI); including higher order corrections for this state could be a fruitful endeavor in future studies. 

As ZnO is usually n-type, we focus on transitions related to the charge change (2-/1-) and (1-/0). The calculated PL value from the (2-/1-) vertical transition is 0.84 eV, a value too small to be detected in prior EPR experiments. The calculated electron capture energy by the neutral Zn vacancy is 1.84 eV, aligning well with the experimental PL results of 1.75 eV \cite{knutsen2012}. Prior work \cite{janotti2009,lyons2017} has pointed out $\text{V}_{\text{Zn}}$ can form a defect complex with hydrogen, resulting in different defect levels. While studies of this complex are outside the scope of this article, this is certainly a topic worthy of further investigation with the approaches presented in this article.

While this work focuses on the applicability of bulk non-empirically tuned hybrid parameters to two native vacancy point defects of ZnO, which exhibit deep levels, shallow defect levels with delocalized character are likely accurately captured with bulk WOT-SRSH parameters as well since such parameters have been shown to be effective for delocalized states in many prior bulk calculations \cite{wing2021,ohad2022,gant2022,ohad2023}. Studying the applicability of bulk WOT-SRSH parameters to shallow defects is an interesting topic for future studies.

\section{Summary}
Focusing on point defects in ZnO, we find that using optimally-tuned SRSH parameters obtained from a pristine bulk cell based on a MLWF (WOT-SRSH), or from a defect-containing supercell based on a defect orbital, can both lead to accurate defect electronic and optical properties. Specifically, we find that the thermal and optical charge transitions of the two native Zn and O vacancies in ZnO calculated using non-empirical optimally-tuned SRSH parameters compare well to experiments and recent empirically tuned theoretical calculations. Our work highlights the potential of WOT-SRSH approach as highly accurate for predicting charge states and spectroscopy of defects in complex materials.

\begin{acknowledgments}
The authors thank Guy Ohad and María Camarasa-Gómez for helpful discussions.
This work is primarily funded by Liquid Sunlight Alliance, a DOE Energy Innovation Hub, supported by the U.S. Department of Energy, Office of Science, Office of Basic Energy Sciences, under Award Number DE-SC0021266. Work on advancing non-empirical tuned hybrid functionals was funded through NSF– Binational Science Foundation Grant No. DMR-2015991 and by the Israel Science Foundation. We acknowledge computational resources provided by the National Energy Research Scientific Computing Center (NERSC), supported by the Office of Science of the Department of Energy operated under Contract No. DE-AC02-05CH11231 using NERSC award BES-ERCAP0024109. This research is also part of the Frontera computing project at the Texas Advanced Computing Center, and used those resources as well. Our work on Frontera is supported by the National Science Foundation award OAC-1818253.
\end{acknowledgments}
\clearpage
\bibliography{main}
\end{document}